\documentclass[aps,prl,showpacs,twocolumn,superscriptaddress,letterpaper]{revtex4-1}
\usepackage{epsfig}
\usepackage{bm}
\usepackage{graphicx}
\usepackage{amssymb}
\usepackage{amsmath}
\usepackage{color}
\usepackage{slashed}
\newcommand{\bfr}{{\bf r}}

\newcommand{\ben}{\begin{displaymath}}
\newcommand{\een}{\end{displaymath}}
\newcommand{\be}{\begin{equation}}
\newcommand{\ee}{\end{equation}}
\newcommand{\bea}{\begin{eqnarray}}
\newcommand{\eea}{\end{eqnarray}}

\newcommand{\eq}[1]{Eq.~(\ref{#1})}

\newcommand{\bfk}{{\bf k}} 
\newcommand{\bfR}{{\bf R}}\newcommand{\bfK}{{\bf K}}

\def\g{\gamma}
\def\e{\epsilon}

\def\a{\alpha}
\def\D{\Delta}\def\d{\delta}

\def\e{\epsilon}
\def\L{\Lambda}

\def\b{\beta}

\def\nno{\nonumber}

\begin{document} 
\title{\bf \hskip10cm \\
{Can Long-Range Nuclear Properties Be Influenced By  Short Range Interactions? A chiral dynamics estimate
} %{Application to nuclear charge radii} 
%\\ --or-- \\
%{the neutron skin thickness} \\
%--or-- \\
%{Ca isotopes}
}
\newcommand*{\UW}{Department of Physics, University of Washington, Seattle, WA 98195}
\newcommand*{\UWindex}{1}
\affiliation{\UW} 
\newcommand*{\MIT}{Massachusetts Institute of Technology, Cambridge,
MA 02139}
\newcommand*{\MITindex}{2}
\affiliation{\MIT}
\newcommand*{\ODU}{Old Dominion University, Norfolk, Virginia 23529}
\newcommand*{\ODUindex}{3}
\affiliation{\ODU}
\newcommand*{\TAU}{Tel Aviv University, Tel Aviv 69978, Israel}
\newcommand*{\TAUindex}{4}
\affiliation{\TAU}
\newcommand*{\NRCN}{Nuclear Research Centre Negev, Beer-Sheva, Israel.}

\author{G.A. Miller}
\email[Contact Author \ ]{miller@uw.edu}
\affiliation{\UW}
\author{A. Beck}
\altaffiliation[On sabbatical leave from ]{\NRCN}
\affiliation{\MIT}
\author{S. May-Tal Beck}
\altaffiliation[On sabbatical leave from ]{\NRCN}
\affiliation{\MIT}
\author{L.B. Weinstein}
\affiliation{\ODU}
\author{E. Piasetzky}
\affiliation{\TAU}
\author{O. Hen}
\affiliation{\MIT}

\date{\today}

\begin{abstract}
Recent experiments and many-body calculations indicate that  approximately 20\% of the nucleons in medium and heavy nuclei ($A\geq12$) are 
part of short-range correlated (SRC) primarily neutron-proton ($np$) pairs. %This is a feature of both symmetric and 
%asymmetric nuclei. % not appropriate for abstract The origin is the short-ranged part of the two-nucleon tensor interaction.
We  find that using  chiral dynamics to account for the formation of $np$ pairs due to the effects of iterated and irreducible  two-pion exchange leads to values  consistent with the 20\% level. 
We  further  apply chiral dynamics to studiy how these correlations  influence   the calculations of nuclear charge radii, that traditionally truncate their effect, to  find that they are  capable of introducing non-negligible effects. %In particular,  and find an increase 
%find that the formation of $np$-SRC pairs between protons and
%For neutrons  outside a closed shell the computed increase found using chiral dynamics  % in an outer shell increases the charge radius. We find that including 
% short-range effects in  calculations of  can lead to significant   underestimates of the mean square 
%charge radius that are  
%is consistent with the
%typical difference between measured charge radii and ab initio 
%calculations that truncate the effects of  SRC pairs. 
\end{abstract}
 
 \maketitle 
\noindent

\section{Introduction}
Electric charge distributions are a fundamental measure of the
arrangement of protons in nuclei~\cite{Angeli:2004kvy}.
%OH nuclear charge distribution. [Or: I don't think it makes sence to say `electric charge dist. measured nuclear charge dist.'. Its 100% true, but trivial]
The variation of charge distributions of elements along isotopic
chains is of particular interest due to its sensitivity to both single particle shell
closure and binding effects, as well as properties of the nucleon-nucleon ($NN$)
interaction. In addition, differences between the neutron and proton
matter radii in neutron rich nuclei are extensively used to constrain the nuclear
symmetry energy and its slope around saturation density, and thus have significant
implications for calculations of various  neutron star properties, including their
equation of state \cite{Hebeler:2010jx,Horowitz:2001ya,Brown:2000pd}. 

Modern charge radius calculations  
 go far
beyond the single particle approximation. Ab initio many-body
calculations can be done using various
techniques~\cite{Barrett:2013nh,Carlson:2014vla,Hergert:2015awm}, but are generally very calculationally
demanding, especially for medium and heavy nuclei. The
recent development of chiral effective field theory  (EFT) inspired soft  interactions that
offer  systematic evaluation of the accuracy of
the calculation, and  methods
%%GM made following a bit more precise
using similarity renormalization group (SRG) evolution techniques to reduce the size of the model space  are especially useful as they significantly simplify
calculations of energy levels. While generally successful, these calculations obtain  a systematic underestimate of nuclear charge radii and their variation along isotopic chains~\cite{Ruiz:2016gne,Simonis:2017dny,Binder:2013xaa}. This is a known, but not yet explained feature these calculations.  

However, both  procedures mentioned above are often implemented in a manner that truncates high momentum can   reduce the high-momentum components of the nuclear wave function 
 that  could be important in computing matrix elements of observable quantities.
Here we discuss the impact of neutron-proton Short-Range
Correlations ($np$-SRC)~\cite{Hen:2016kwk,Hen:2014nza,Subedi:2008zz} on nuclear charge radii. 
%This study was motivated by a recent
%measurement ~\cite{Ruiz:2016gne} of the charge radii of 
%i%sotopes from $^{39}$Ca to $^{52}$Ca that found an
%unexpectedly large increase of about $\delta \langle r^2\rangle \approx 0.53$ fm$^2$
%in the mean square (MS) charge radii from $^{48}$Ca to $^{52}$Ca.
%This measured increase was 0.1 to 0.3 fm$^2$ greater than
%that obtained using various ab initio calculations, primarily using soft $NN$ interactions.
%that truncate short-range physics. 
%Systematic underestimation of the charge radii of 
%nuclei is a known, but   unexplained, global feature of such ab initio 
%calculations~\cite{Simonis:2017dny,Binder:2013xaa}.
%not relevant here this is a discussion point,
%Garcia Ruiz {\it
%et al.} speculated that this difference could be due to a lack of
%``deformed intruder states associated with complex configurations'' in
%the models~\cite{Ruiz:2016gne}.

Over the last decade considerable evidence has accumulated regarding
the attractive nature of the short-ranged neutron-proton interaction
in the spin triplet channel. Measurements of  relatively high-momentum transfer  inclusive electron-scattering eactions   indicate that about 20\% of the
nucleons in medium and heavy nuclei ($A\geq12$) have momentum greater
than the nuclear Fermi momentum ($k_F\approx $ 275 MeV/c)~\cite{frankfurt93,egiyan03,egiyan06,fomin12,Duer:2018sxh}. 
In the momentum range of 300 -- 600 MeV/c, these high-momentum nucleons were observed to be predominantly 
members of $np$-SRC pairs, defined by having large relative and smaller center-of-mass momenta relative to $k_F$ ~\cite{tang03,piasetzky06,shneor07,subedi08,korover14,Hen:2016kwk,Hen:2014nza,Cohen:2018gzh}. This is an operational 
momentum-space definition of the term `short-range'. 

Calculations indicate that the origin of these correlated $np$-SRC pairs lies in 
the action of a strong short-ranged tensor interaction~\cite{Alvioli:2007zz,Schiavilla:2006xx,Sargsian:2005ru}. These experiments and interpretive calculations are based on the idea that at high-momentum transfers the reaction factorizes, allowing cancellation of reaction mechnism effects in cross-section ratios of different nuclei and the use of the impulse approximation (with suitable modest corrections) in other cases~\cite{Hen:2016kwk,Frankfurt:2008zv}. In effective field theory language, this corresponds to using a high-resolution scale in the similarity renormalization group SRG 
transformations~\cite{Furnstahl:2013oba}.

In the case of neutron-rich nuclei, recent measurements \cite{Duer:2018sby} indicate 
that, from Al to Pb, the fraction of correlated neutrons (i.e., the
probability for a neutron to belong to an $np$-SRC pair) is  approximately constant while 
that of the protons grows with nuclear asymmetry approximately as $N/Z$ (where $N$ and $Z$ 
are the numbers of neutrons and protons respectively). This is a first indication of 
a possible significant impact of $np$-SRC on the proton distributions in neutron
rich nuclei.

The effects of tensor-induced $np$-SRC pairs in nuclei were shown to 
have significant impact on issues such as the internal structure of bound
nucleons (the EMC effect) \cite{weinstein11,Hen:2012fm,Hen:2013oha,Hen:2016kwk},  
neutron-star structure and the nuclear symmetry energy at supra-nuclear densities \cite{Frankfurt:2008zv, hen15, Cai:2015xga,Hen:2016ysx}, 
the isospin dependence of nuclear correlation functions \cite{Cruz-Torres:2017sjy}.
However,
it is natural to expect 
that due to the tensor interaction's short-ranged nature, its impact
on long range (i.e., low-energy) observables can be neglected. 

This expectation is examined here by focussing on 
the  particular problem highlighted in Ref.~\cite{Ruiz:2016gne}: the computed difference in charge radii between $^{52}$Ca and $^{48}$ Ca is smaller than the measured value. The ab initio calculations in that work are based on several interactions that include NNLO$_{\rm sat}$, which is fit to scattering data up to 35 MeV and to certain nuclear data from nuclei up to A=24~\cite{Ekstrom:2015rta}, and the interactions SRG1 and SRG2. The latter are derived from the nucleon-nucleon interaction of Ref.~\cite{Entem:2003ft} by performing an evolution to lower resolution scales via the similarity renormalization group (SRG)~\cite{Furnstahl:2013oba}. 
 Ref.~\cite{Ruiz:2016gne}  speculates that  the reason for the discrepancy between theory and experiment is a  lack  in the description of deformed intruder states associated with complex configurations.  The  NNLO$_{\rm sat}$ interaction does give the correct $^{40}$Ca radius and includes short-range correlations for like nucleon pairs implicitly through its optimization procedure.

Here we study another possible reason for the discrepancy that stems from the use of soft interactions.
We view all of the interactions employed in  Ref.~\cite{Ruiz:2016gne} as soft interactions that reduce the influence of short-range correlations.  We argue  that including omitted effects of  $np$-SRC pairs between protons and  neutrons 
in the outer shells of neutron-rich nuclei may change the computed value of the proton MS charge radius 
 for some neutron-rich nuclei. Note that the use of soft interactions, caused by using  form factors that reduce the probability for high-momentum transfer should be accompanied by including  modified electromagnetic currents  as  demanded by current conservation. If  instead the SRG is used, it should be accompanied by corresponding unitary transformations on the operators. If these modifications to current operators are  done completely accurate calculations may be  possible.

We begin by using  the simplest possible  illustration of  the possible impact of SRCs on computed charge radii.
This is based on  a   two-state
system and is intended only   to explain the basic idea. We then use chiral dynamics to estimate the probability of short-range correlations, and  further study how these correlations  may 
influence calculations of nuclear charge radii. 
%Next, we
%use a simple model with a short-range attractive neutron-proton interaction to calculate 
%the increase in the mean-square charge radius 
%when one neutron is added to $^{48}$Ca. 

\section{Schematic model} %Possible Impact of high-momentum states} 
Consider the evaluation of an operator ${\cal O}$ in a framework
in which SRC effects can be  truncated. We examine the effect of this truncation on the computation of relevant matrix elements.
%we can evaluate the general impact of SRCs on the relevant operators.
 % We   focus on the calculation of the proton MS charge radius and  use the language of the  SRG procedure in the beginning of this section.

The consistent application~\cite{Schuster:2013sda,Neff:2015xda} 
of SRG evolution in a many-body
calculation requires that the Hamiltonian, $H$, as well as all other
operators, be transformed according to ${\cal O} \rightarrow U {\cal
O}U^\dagger .$ Here  $U$ is a unitary operator, chosen to simplify the evaluation of energies
by reducing the matrix elements of $H$ between low- and high- momentum subspaces. The aim is to obtain a block-diagonal Hamiltonian. 
 Such transformations  have 
no impact on observables and include the effects of short-range physics. 
However, in the case of proton MS charge radius calculations, several works~\cite{Ruiz:2016gne} 
further simplify the calculation by evaluating the expectation value of the {\it un-transformed} 
mean square charge radius operator on the {\it transformed} wave functions.

To understand the general effect of using such un-transformed operators  we consider a simple two-state model with two components, 
$|P\rangle$ and $|Q\rangle$, respectively representing low-lying shell model states
and high-lying states within the model space. The $Q$-space  is intended to 
represent the states that enter into the many-body wave function due to the 
short-range correlations. Thus the $Q$-space  dominates the high-momentum part of the ground 
state one-body density~\cite{Weiss:2016obx}. 

Since we are concerned with nuclear charge radii,
the simple model must be further defined by the matrix elements of the
charge radius squared operator, $R^2$.
This operator is of long range and is not expected to allow much connection between the $P$ and $Q$ spaces. Therefore, this model is defined by the simple statement: $\langle P|R^2|Q\rangle=0$. The motivation for this statement comes from the single-particle  harmonic oscillator model: the action of the square of the radius changes the principal quantum number by at most one unit.  One may arrange model spaces satisfying this condition by using superpositions of harmonic oscillator wave functions. A consequence of this is that   $\langle Q|R^2|Q\rangle-\langle P|R^2|P\rangle>0$.
%It follows that $\langle Q|R^2|Q\rangle-\langle P|R^2|P\rangle\ge 3/(M\Omega)$ where $M$ is the proton mass.

The Hamiltonian   for a two-state system is given by
\bea
H=\left[
\begin{matrix}
-\e & V \\
V & \e 
\end{matrix}
\right]
\eea
where $2\e$ is the energy splitting between $|P\rangle$ and $|Q\rangle$
and $V$ is the short-distance coupling between them.
The exact ground state $|GS\rangle$ can be computed and the occupation probability of the $Q$ space is 
given by ${\cal P}_Q\equiv V^2/[(\epsilon+\D)^2+V^2]$, with $\D \equiv
\sqrt{\e^2+V^2}$. ${\cal P}_Q$ in this 
model corresponds to the probability that a proton belongs to a short
range correlated pair. %, ${\cal P}_Q = {\cal P}_{SRC}$. 
The mean-square charge radius is then given by:
\bea&
\langle GS|R^2|GS\rangle=%\nonumber\\%\langle P |R^2|P \rangle {(\e+\D)^2\over (\e+\D)^2+V^2}+\langle Q |R^2|Q \rangle {V^2\over (\e+\D)^2+V^2}\\&=
\langle P |R^2|P \rangle+(\langle Q |R^2|Q \rangle - \langle P
|R^2|P \rangle){\cal P}_Q \nno\\&
\label{exact}
\eea
with the second term representing the influence of the high-lying states.

%% GM Following sentence is redundant with next para
%Next we explain how this effect is truncated in many SRG calculations and then perform a quantitative estimate of its magnitude.

% and then turn to a quantitative estimation of the
%magnitude of this enhancement for the case of Ca isotopes.

%Utilizing the fact that, for an harmonic oscillator basis, the diagonal elements of the kinetic and potential energies 
%high-energy harmonic oscillator states have large radii so that the second term 
%in \eq{exact} is always positive. We will show this explicitly for a specific case below.

%In this harmonic oscillator basis, the diagonal elements of the kinetic and potential energies are equal.
%Thus, a high energy state has large momentum and large radius. 
%The equality between kinetic and potential energies means that
%\bea \langle Q|R^2|Q\rangle>\langle P|R^2|P\rangle,\label{qrp}\eea
%which is valid for any attractive energy-independent single-particle potential of finite range. 
%Thus, the effect of SRCs is guaranteed to increase the mean square charge radius obtained in the present simple model.

%One should note that for sufficiently large energies, the eigenstates are in the continuum and the average MS radius is infinite. 
%This is resolved by understanding the role of single-particle states in the two-nucleon correlation function, as explained in the next section.

We interpret the result \eq{exact} in terms of the SRG.
For a two-state system the complete SRG
transformation simply amounts to diagonalizing the Hamiltonian. The
resulting renormalization-group improved Hamiltonian $\tilde{H}$ is a
diagonal matrix with elements given by $\pm \D$. The unitary transformation sets the matrix elements of $\tilde H$ between the low and high momentum sub-spaces to zero.
Applying the same unitary transformation to the
operator $R^2$, one would obtain the same result as
\eq{exact}. However, the procedure of Ref.~\cite{Ruiz:2016gne}, and
 others, for example, ~\cite{Simonis:2017dny,Binder:2013xaa}, corresponds to simply using
\bea \langle GS|UR^2U^\dagger|GS\rangle\approx
\langle GS|R^2|GS\rangle=\langle P |R^2|P \rangle.\label{ng}
\eea
which contrasts with the complete calculation of \eq{exact}. For the given model, $(\langle Q |R^2|Q \rangle - \langle P
|R^2|P \rangle)>0$, so that the omission of the unitary transformation on the $R^2$ operator leads to a reduction in the matrix element. In realistic situations the correction term $(\langle Q |R^2|Q \rangle - \langle P
|R^2|P \rangle){\cal P}_Q $ would be  the difference between two large numbers, so it could be positive or negative.

 %We show below that using \eq{ng} leads to an underprediction of the charge radius for neutron rich nuclei.
%For neutron-rich nuclei where the extra neutrons are in higher
%single-particle states with larger radii, this second term may  lead to an increase in the proton
%MS radius, due to the attractive tensor interaction between protons with neutrons in the outer parts of the nucleus.

The two-component model presented above shows the possible qualitative effect
on the computed mean-square radius of truncating SRCs. However, it
cannot be used to make a quantitative prediction. 

\section{Chiral Dynamics Estimate}

We now turn to a more complete calculation  to estimate the magnitude of the effect for 
  the specific case of adding a neutron to the $1f_{5/2}-2p_{3/2}$ shell around a $^{48}$Ca core.
The additional neutron is mainly located in the outer edge of the nucleus.
In the pure shell model, this neutron would not affect the proton MS charge radius.
However, it is natural to wonder if the attractive, short-range neutron-proton tensor
interaction that creates $np$-SRC pairs will cause the protons to move
closer to the additional neutron, thereby increasing the charge radius.

The calculation starts with a $^{48}$Ca wave function that has been
obtained using an interaction that  explicitly includes the effects of high-momentum components, and considers the 
effect of a missing short-ranged potential $V$ on an $np$ product wave function $|n,\a)$,
where $n$ and $\a$ represent the neutron and proton orbitals respectively. For numerical work, in this exploratory effort,  we ignore the spin orbit force and use harmonic oscillator single-nucleon  wave functions.
 with frequency, $\Omega=10.3$ MeV, that yields the 
measured charge radius of 3.48 fm, and the corresponding length parameter, $b^2=1/(M\Omega)=4 .02\,{\rm fm}^{2}$.
The interaction $V$ is meant to represent the effects including the short-range strength masked by the SRG procedure, and also corrections to the effects of using a very soft nucleon-nucleon interaction.
The effect of $V$ on the wave function 
is given by
\bea %\widetilde
{ |n,\a\rangle}=C^{-1/2}\left[ |n,\a)%\rangle
+ {1\over \D E}{QG} |n,\a) \right],\label{Gmat}
\eea
where $G$ is the Bruckner $G$-matrix that sums the interactions $V$ on
the pair and includes the effects of short-range correlations. Note
that $|\rangle$ represents the full wave function while $|)$ represents the product state. The energy denominator $\D E \equiv E_{n}+E_{\a}-H_0$, with $H_0$ the Hamiltonian in the absence of $V$, $Q$ the projection operator that places the neutron and proton above the Fermi sea and 
$C_{n\a}$ is the normalization constant:
\bea
&C_{n\a}=%1+\langle n,\a|GQ{1\over \D E} {1\over \D E}QG |n,\a\rangle\nonumber\\&=
%1+\sum_{m,\b>E_F}{\langle n,\a|G|m\b\rangle\langle m,\b| G |n,\a\rangle \over (E_{n}+E_{\a}-E_{m}-E_{\b})^2}
1+I_{n\a},
\eea
with $I_{n\a}\equiv \sum_{m,\b>E_F}{( n,\a|G|m\b)( m,\b| G |n,\a) \over (E_{n}+E_{\a}-E_{m}-E_{\b})^2}.$
%{\bf Yes this is $C$ and not $C^{-1}$}
%Defining the second term on the right-hand-side as $I_{n\a}$ 
%The probability of the short-range correlation is
%\bea
% $ P^{SRC}_{n\a}={I_{n\a}\over 1+I_{n\a}}
%.$ %\eea
%Many calculations over a long period of time have found that $P^{SRC}_{n\a}$ is about 0.2. {\bf Or has data showing this is independent of $n,\a$.}

%Now turn to calculating the charge radius.
%In the following we neglect the small effects of the neutron charge density
%and the non-zero extent of the proton so that the entire nuclear
%charge density is carried by point protons. 
Defining the proton MS radius as the expectation value of the
operator $R^2_p$, 
%Then the nuclear ch
%mean-square charge radius is obtained by computing the expectation
%value of the square of the proton's nuclear radius 
we want to know the quantity
\bea & { \langle n,\a|} R^2_p { |n,\a\rangle}=\nno\\&C_{n\a}^{-1}[ (n,\a| R^2_p |n,\a)+ %\langle 
(n,\a|{GQ}{1\over \D E} R^2_p{1\over \D E} QG|n,\a)] ,\nno\\&\label{full}
\eea
obtained by using the fact that the one-body operator $R_p^2$ does not connect the states $|n\a)$ and $|m\b)$ that differ by two orbitals. % if $m\ne n$. %The second term is positive definite. 
Using the definitions:
\bea 
&R^2_{n\a}\equiv {1 \over I_{n\a}}( n,\a|{GQ}{1\over \D E}
R^2_p{1\over \D E} QG|n,\a) 
\eea
and
\bea
R^2_\a\equiv (\a|R_p^2|\a),
\eea
one finds
\bea { \langle n,\a|} R^2_p { |n,\a\rangle}=R^2_\a+ {\cal P}^{SRC}_{n\a}(R^2_{n\a}-R^2_\a), \label{good1}
\eea
with  ${\cal P}^{SRC}_{n\a}={I_{n\a}\over 1+I_{n\a}} \approx 0.2$ is
the measured SRC probability~\cite{frankfurt93,egiyan03,egiyan06,fomin12}. Probabilities of this size have been obtained by both ancient and modern  computations
~\cite{Davies:1971zg,Davies:1974zz,Hasan:2003eh,Carlson:2014vla,Hu:2017src,Wiringa:2013ala,Lonardoni:2017egu}.
The effect of SRCs is embodied in the second term of \eq{good1}, as in \eq{exact}. Note the similarity between the two equations. Given that  ${\cal P}^{SRC}_{n\a}$ cannot be zero,  the effect of SRCs must  either increase or decrease the
computed mean-square charge radius. In some cases, there could be a cancellation between the two parts of the correction term, but this should not be expected to occur for all configurations.

We  next show  that the $\sim$ 20\% value of ${\cal P}^{SRC}_{n\a}$ is consistent with the chiral dynamics treatment of Refs.~\cite{Kaiser:1997mw,Kaiser:2001jx}. In that work
 the chiral dynamics  interaction, dominated by the iterated effects of the one pion exchange potential (OPEP) is shown to be the cause of nuclear binding. Furthermore, the reduction of this effect with increasing nuclear density through the effects of the Pauli principle on the intermediate nucleon-nucleon state is responsible for nuclear saturation. The ideas of Refs.~\cite{Kaiser:1997mw,Kaiser:2001jx} account for the qualitative features of nuclear physics with startling simplicity. 
 
The key point for us is that the  dominant source of attraction comes from a short distance effect, well-represented by zero-range delta function interaction  (denoted  here as $V_{00}$) proportional to a cut-off parameter $\L$, that is consistent  with the requirements of chiral symmetry. This interaction accounts for an attraction ``in the  hundred MeV range (per nucleon) for physically reasonable values of the cut-off, $0.5<\L<1.0$ GeV."  These authors also note  that the two pion exchange interaction contains a  factor that favors the isospin 0 two-nucleon state by a factor of 9 over the isospin 1 state. The $np$ dominance discussed above is a natural consequence of the chiral dynamics interaction.

%  Here we consider the situation in
%which the neutron occupying the orbital $n$ has a large probability to
%be at the edge of the nucleus and the proton occupying the orbital
%$\a$ is mainly located in the interior. Then we expect that
%$R^2_{n\a}-R^2_\a>0$ and thus including the influence of the SRCs increases
%the proton MS radius.

%We evaluate the possible changes in computed charge radii by  using  the chiral dynamics  interaction, dominated by the iterated effects of the one pion exchange potential (OPEP), presented in Refs.~\cite{Kaiser:1997mw,Kaiser:2001jx}. This iteration effect is %seen to be the cause of nuclear binding. Furthermore, the reduction of this effect with increasing nuclear density through the effects of the Pauli principle on the intermediate nucleon-nucleon state is responsible for nuclear saturation. The attraction comes from a %short distance effect, well-represented by zero-range delta function interaction,  denoted  as $V_{00}$, proportional to a cut-off parameter $\L$, that is not in conflict with the requirements of chiral symmetry. This interaction accounts for an attraction ``in the  %hundred MeV range (per nucleon) for physically reasonable values of the cut-off, $0.5<\L<1.0$ GeV." 

The calculation of the iterated OPEP was repeated 
 in Appendix A of
Ref.~\cite{Hen:2016kwk}. In agreement with \cite{Kaiser:1997mw,Kaiser:2001jx},  that work  shows that the iteration of
the spin-triplet one-pion exchange (OPE)  potential (including the
transitions $^3S \rightarrow{}^3D\to{}^3S $) results in a strong attractive
effective short-range interaction that acts in the relative
$S$ state (see also \cite{Colle:2013nna,Colle:2015ena}).   The strength is of the correct magnitude to roughly account for
the deuteron $D$-state probability, and therefore may be expected to
roughly account for the probability that an $np$ pair is in a short range
correlation.

Refs.~\cite{Kaiser:1997mw,Kaiser:2001jx} explains that $V_{00}$ should {\it not}  be iterated with itself or with $1\pi$ exchange. Therefore the iterated OPE  term $V_{00}$
approximately corresponds to the  $G-$matrix of \eq{Gmat}.
That reference gives
\bea & V_{00}(\bfr)=-{8\pi^2 }M \L({g_A\over 4\pi f_\pi})^4(1-{(3g_A^2+1)(g_A^2-1)\over 10 g_A^4})\nonumber\\&\times(1-2.18{k_F\over M})(3-2\tau_1\cdot\tau_2)\d(\bfr)\nonumber\\&\equiv V_0(3-2\tau_1\cdot\tau_2)\d(\bfr),\eea
where $M$ is the nucleon mass, $g_A$ the axial vector coupling constant, and the pion decay constant $f_\pi=$ 92.4
 MeV. The effect of the Pauli principle is encoded in the term proportional to $k_F$. This expression includes the smaller repulsive effects of the irreducible two-pion exchange graphs. 
 
 The interaction causes a change in the wave function:
\bea |\d \Phi_{n,\a})\equiv{Q\over \D E }V_{00}|n,\a),\label{dp}
\eea
which accounts for the second term of \eq{Gmat}.
To proceed, we need the operator ${Q\over \D E }$. We use its value in nuclear matter under the reference spectrum approximation~\cite{Day:1967zza},  which is qualitatively valid for use along with short-ranged interactions~\cite{Brown:1967}. Then in the  coordinate-space representation ($\bfr_{1,2}=\bfR \pm \bfr/2$) this is  approximated by  
the expression:
\bea &
\langle \bfR,\bfr| {Q\over \D E }|\bfR',\bfr'\rangle=-\int{d^3K \over (2\pi)^3}{d^3k\over (2\pi)^3}\times\nonumber\\&{e^{i\bfK\cdot(\bfR-\bfR')}e^{i\bfk \cdot(\bfr-\bfr')}\over 2A_2+W  +{K^2\over 4M^*} +{k^2\over M^*}},
\eea
where $M^*=0.6M, A_2\approx 100 $ MeV, and $W$ is a starting energy~\cite{Day:1967zza}.  The quantity  $k_F$ is the effective Fermi momentum for a given nucleus, here taken as 1.36  fm$^{-1}$. The short-ranged nature of the interaction is expected to excite  that part of the phase space that  involves regions of relative low $K$ and relatively high $k$. Thus, the integral is simplified by replacing $K^2$ by its average value of $(6/5) k_F^2/M$, where    $k_F$ is the effective Fermi momentum for a given nucleus. Then defining $\g^2\equiv M^*( 2A_2+W +(3/10) k_F^2$, we find that
\bea &
\langle \bfR,\bfr| {Q\over \D E }|\bfR',\bfr'\rangle\approx \delta(\bfR-\bfR')f(|\bfr'-\bfr|),
\eea
with $
f(r) = {M^*\over 4\pi r} e^{-\g r}.$ %\equiv M\int {d^3k\over(2\pi)^3}{e^{i\bfk\cdot\bfr}\Theta(k^2-0.7k_F^2)\over \bar B M +k^2}.
 Then 
\bea \langle \bfR,\bfr| \d\Phi_{n,\a})\approx f(r)V_0(3-2\tau_1\cdot\tau_2)\phi_n(\bfR)\phi_\a(\bfR). \eea
Then 
\bea &\sum_\a'I_{n\a}\approx \sum_\a'\int d^3R |\phi_n(\bfR)\phi_\a(\bfR)|^2 41 {(M^*V_0)^2\over 8\pi\g}
%I%(x_0)\equiv \int_{x_0}^\infty dx {x^2\over (1+x^2)^2},
\eea 
where $\sum_\a'$ denotes an average over the bound proton levels. The factor 41 arises from the average of the contributions of $9^2=81$ (from the $T=0$ component of an $np$ pair) and 1 (rom the $T=1$ component).    %is value is  roughly  consistent with the expectations of measurements and previous calculations of ${\cal P}^{SRC}$ discussed above.
Given the change in wave function of \eq{dp} 
 one finds  the result that
\bea
R^2_{n\a}\approx {\int d^3R R^2 |\phi_n(R)\phi_\a(R)|^2 \over \int d^3R |\phi_n(R)\phi_\a(R)|^2}. %+{1\over 16 M \bar B}.\label{qual}
\eea
%%GM
This  amounts to using the known Zero-Range Approximation, previously shown to successfully reproduce SRC effects~\cite{Colle:2015ena}. 
%To see how $R^2_{n\a}>R^2_\a$ may occur, consider the simple example in which $|\phi_n(R)|^2$ is peaked sharply at the surface position $R_S^2>R_\a^2$. Then $R^2_{n\a}-R_\a^2\approx R_S^2-R^2_\a>0.$ This illustrates the 
%idea that a neutron at the surface may  attract an inner proton outwards.

We next apply this idea to the specific case of adding neutrons to $^{48}$Ca, with the
added  neutron in either  the $n=0f_{5/2}$ or $n=1p_{3/2}$ state.  We will use  the average over the $L_z$ substates.
% we compute the contributions from the occupied proton orbitals. 
Given parameters mentioned above, we find that \bea {\cal P}^{SRC}_{n\a}\approx 0.2\pm 0.02 \eea for all of the orbitals considered. The dependence on the starting energy $W$ is weak because it is much smaller than $2A_2=200 $ MeV.
%%GM solv 
Then the  effects of short-range correlations on  the mean-square charge radius of an orbital $n$  are given by:
\bea
\D R^2(n)  \equiv\sum_\a'{\cal P}^{SRC}_{n\a}(R^2_{n\a}-R^2_\a) % P_{SRC}* {0.19 b^2}=
%\D R_{ch}^2= P_{SRC} ({3b^2})= 2.4\,{\rm fm^2}. 
\eea
Numerical evaluation leads to
\bea \D R^2(0f_{5/2})=0.17 \pm0.12\, {\rm fm}^2,\,\,
 \D R^2(1p_{3/2})=-0.8\,  {\rm fm}^2.\nonumber\\
 \eea
 The variation seen for $\D R^2(0f_{5/2})$ results from increasing the $b$-parameter by an amount up to 10 \%. This is done because this  state is weakly (if at all) bound. The value of  $ \D R^2(1p_{3/2})$
 is relatively insensitive to such changes. Its negative nature arises from the node in $1p_{3/2}$ state wave function which appears in the nuclear surface. This calculation  may appear somewhat crude, but is reasonably well-constrained by using  ${\cal P}^{SRC}_{n\a}\approx0.2$ and  the measured charge radius of the $^{48}$Ca core state. %We cannot say that the effect we examine helps in solving the problem of under computing nuclear charge radii. However, t
 The effect under consideration  is not obviously zero, and 
the specific values we find are  
large enough to be taken seriously. Thus we   suggest that the effect of short range correlations on computed charge radii should be considered in more detailed calculations that include many configurations and aim at high precision.

We note  that the ab initio triton  calculation of~\cite{Schuster:2014lga} finds that ignoring the effect of the unitary transformation on the charge radius squared operator leads to an   overestimate of the value by about 0.04 fm , not dissimilar in percentage to our results.  

%The actual nuclear charge radius will be sensitive to many effects in
%addition to the neutron-proton tensor interaction described above.
%For example, as neutrons are added to $^{40}$Ca, the extra neutrons 
%form short-range correlated pairs with the protons, increasing the
%fraction of protons in correlations. The measured charge radius
%initially increases because the $1f_{7/2}$ neutrons are at larger
%radius than the average proton and the correlations increase the
%radius of each proton shell. When more than four neutrons are added to
%$^{40}$Ca, the charge radius decreases as
%the extra neutrons increase the nuclear binding, making the entire
%nucleus more compact. This leads to the same measured proton MS
%radius for $^{40}$Ca and $^{48}$Ca, even though the latter has higher binding and proton separation energies. 
%Therefore, the specific calculation for $^{49}$Ca described above is only practical near the closed-shells of nuclei, where mean-field 
%wave-functions can be used and where adding a few neutrons does not 
%significantly change the overall binding. 

% However, since the proton SRC correlation probability increases as a function of neutron
% excess \cite{duer17}, we expect the increase in the proton MS charge
% radius due to $np$-SRC pairs between the protons and the outer-shell neutrons 
% to be a generic feature of neutron rich nuclei.

\section{Discussion}

Ref.~\cite{Ruiz:2016gne} explains various possible reasons (in addition to the effects of deformed  intruder states)  for the differences  between the experimental  and computed values of charge radii. 
  The  calculations are based on the single-reference coupled-cluster method, which is ideally suited for nuclei with at most one or two nucleons outside a
closed (sub-) shell. Many calculations in the literature considerably underestimate the large charge radius of $^{52}$Ca.    Dynamical quadrupole and octupole effects  could be responsible,  but seem to be too small. These authors mention core breaking effects. Their {\it ab iniitio} calculations  do show ``a weak, but gradual erosion of the proton core as neutrons are added."    Thus our statement is that   including the effects of short range correlations could make the erosion of the proton core stronger and influence the resulting computed charge radii. Doing the necessary {\it ab initio} calculation is beyond the scope of this paper, but   we   do suggest that   such effects should be considered in more detailed calculations of charge radii that aim at high precision.

We next  discuss the motivation for pursuing such calculations.
The experimental data accumulated in the last decade, in conjunction
with un-truncated ab initio and effective calculations, allows quantifying the abundance
and properties of SRC pairs in nuclei with unprecedented detail. Recent measurements
of asymmetric nuclei indicate that SRC pairs are dominated by $np$ pairs even in very
neutron rich nuclei \cite{Duer:2018sby}. This implies that the protons are more correlated than neutrons
in neutron rich nuclei.

Modern calculations relate the thickness of the neutron skin of
$^{208}$Pb to the nuclear equation of state and hence to neutron star
properties \cite{Hebeler:2010jx,Horowitz:2001ya,Brown:2000pd}. 
A high priority experiment at
Jefferson Lab is planned to measure the neutron skin thickness using parity
violating electron scattering \cite{Horowitz:2013wha}. However, hitherto
neglected short-range
effects which reduce the neutron skin thickness should be
included to accurately relate the neutron skin thickness to the
nuclear EOS.

The indicated larger probability of protons to be part of SRC pairs in neutron
rich nuclei, in combination with the larger average neutron radius,
indicates that short-range correlations can have an impact on 
long-range nuclear properties such as the nuclear MS charge radius.
 The 
present work shows that the 20 \% probability for an $np$ pair to strongly correlated is a natural consequence of 
the chiral dynamics of Refs.~\cite{Kaiser:1997mw,Kaiser:2001jx}. This effect is dominated by the tensor force which is large in the state with deuteron quantum numbers.
 Moreover, we find that
 including the effects of $np$-SRCs   seems to be necessary to achieve a calculation of high precision. This is somewhat surprising because 
 it is a long-range effect of a short-range, 
typically truncated, part of the $NN$ interaction.  
 We hope that the idea  presented here can be confirmed or ruled out by 
more advanced computations.

%The considerations discussed here indicate that 
%the influence of tensor induced short-range correlations between protons and
%neutrons in neutron rich nuclei increases the proton charge radius and
%decreases the neutron matter radius. 
%This indicates that accurate calculations of the proton and neutron MS charge and matter radii need to 
%include the effects of $np$-SRC pairs. If transformations are used to
%simplify the Hamiltonian, the same transformations must be used on
%the operators used to calculate observable properties. Similarly, the use of overly soft nucleon-nucleon interactions could lead to inaccurately computed charge radii.

%%% I comment out the following because carrying out the calculations is not our task in this paper
% in a direct and/or effective manner. The can be done in various
%ways, including calculations using un-truncated interactions, consistent evolution of the charge radius operator, fitting interaction parameters to heavy nuclei data, consistent operation of correlation operators on uncorrelated wave functions and more.
\section*{Acknowledgements}

G.A. Miller would like to thank the Lab for Nuclear Science at MIT for support that enabled this work. We thank A. Schwenk, R. Stroberg and U. van Kolck for useful discussions. This work was supported by the U. S. Department of Energy Office of Science, Office of Nuclear Physics under Award Numbers DE-FG02-97ER-41014, DE-FG02-94ER40818 and DE-FG02-96ER-40960, the Pazy foundation, and by the Israel Science Foundation (Israel) under Grants Nos. 136/12 and 1334/16.

\bibliography{references,eep,emc}

\end{document}